Unconventional superconductivity in low density electron systems and conventional superconductivity in hydrogen metallic alloys.

M.Yu. Kagan

P.L. Kapitza Institute for Physical Problems, Russian Academy of Sciences, Moscow 119334, Russia
National Research University Higher School of Economics, Moscow 109028, Russia
kagan@kapitza.ras.ru

22 April 2016

In a short review-article we first discuss the results, which are mainly devoted to the generalizations of the famous Kohn-Luttinger mechanism of superconductivity in purely repulsive fermion systems at low electron densities. In the context of repulsive-$U$ Hubbard model and Shubin-Vonsovsky model we consider briefly the superconducting phase diagrams and the symmetries of the order parameter in novel strongly correlated electron systems including idealized monolayer and bilayer graphene. We stress that purely repulsive fermion systems are mainly the subject of unconventional low-temperature superconductivity. To get the high temperature superconductivity in cuprates (with $T_C$ of the order of 100 K) we should proceed to the t-J model with the Van der Waals interaction potential and the competition between short-range repulsion and long-range attraction. Finally we stress that to describe superconductivity in metallic hydrogen alloys under pressure (with $T_C$ of the order of 200 K) it is reasonable to reexamine more conventional mechanisms connected with electron-phonon interaction. These mechanisms arise in the attractive-$U$ Hubbard model with static onsite or intersite attractive potential or in more realistic theories (which include retardation effects) such as Migdal-Eliashberg strong coupling theory or even Fermi-Bose mixture theory of Alexandrov, Ranninger and its generalizations.

I) Introduction

Recent discovery of Cooper pairing at 190 K in metallic hydrogen sulfide $H_2S$ under high pressure revives our hopes to move forward from high-temperature to room-temperature superconductivity [1-5]. In the same time there are many interesting low-temperature superconducting systems with anomalous types of pairing and a nontrivial structure of the order parameter.

In the beginning of the review-article we discuss the mechanisms of unconventional superconductivity in 3D and 2D fermionic systems with purely repulsive interaction at low densities. We briefly discuss the phase-diagrams of these systems in free space and on the lattice and find the areas of the superconductive state in the framework of the Fermi-gas model with hard core repulsion, the Hubbard model [6] and the Shubin-Vonsovsky model [7,8]. We demonstrate that the critical superconductive temperature can be greatly increased in the spin-polarized case or in a two-band situation already at low densities.

The proposed theory is based on the Kohn-Luttinger mechanism of superconductivity [9] and its generalizations and explains or predicts p-, d-, f- and anomalous s-wave pairing in various materials such as the idealized monolayer and bilayer graphene, heavy-fermion systems, layered and organic superconductors, iron-based superconductors, superfluid He-3 and spin-polarized He-3–He-4 mixtures, imbalanced 3D and 2D Fermi-gases in the restricted geometry of magnetic traps [10-19].

To get really high superconductive temperatures we should either consider the repulsive models at high electron densities (for instance in parquet situation close to Van Hove singularities [20-22]) or not the purely repulsive models (like the t-J model for cuprates with $T_C$ = 100 K and the Van der Waals interaction potential consisting of strong Hubbard repulsion on one site and weak antiferromagnetic attraction on neighboring sites [23, 24]).

If we change the sign of the interaction completely and consider an attractive-$U$ Hubbard model [25, 26], then at low electron density and strong attraction, the high critical temperature is connected with bosonic character of superconductivity [27-29]. Here at higher temperatures we get local pairs formation in real space at the Saha crossover temperature $T^*$ [27, 30] and then their Bose-Einstein condensation at the lower critical temperature $T_C$ (as in the problem of BCS-BEC crossover [27-30] for ultracold Fermi-gases in the regime of Feshbach resonance [31-39]). In this case we have a new state of matter at intermediate temperatures, namely the normal bosonic metal with interesting transport and thermodynamic characteristics, including very nontrivial square root type ($R \sim \sqrt{T}$) temperature behaviour of the resistivity in the 2D case [25, 26].

At higher electron densities in this class of models the two-component Fermi-Bose mixture of local pairs and one-particle excitations arises [40-45] and defines superconductive and transport properties of the system (as in the case of $BaKBiO_3$ superconducting oxides).

We should stress that unconventional mechanisms compete for different materials with electron-phonon interaction in more conventional weak-coupling (BCS) [46-50] or Migdal-Eliashberg strong-coupling superconductive systems [51-56]. This situation is likely to occur in metallic hydrogen sulfides and possibly in pure metallic hydrogen under very high pressures of several Megabars [1-5, 57-71].

II) Kohn-Luttinger effect in repulsive Fermi-gas model

It was widely accepted that fermion systems with purely repulsive interaction remain in the normal state at very low temperatures. Challenging these naive expectations Kohn and Luttinger in 1965 [9] proposed a new mechanism of superconductivity in repulsive fermion systems based on the presence of Kohn's anomaly [72] $(q-2p_F)\ln|q-2p_F|$ or Friedel oscillations [73, 74] $\cos(2p_F r)/(2p_F r)^3$ in the effective interaction $U_{\text{eff}}$ of two femions via polarization of the fermionic background in 3D. Unfortunately they predicted very low critical temperatures of d-wave pairing for superfluid transition in He-3 and superconductive transition in electron plasma of metals and that was a reason why their results were unjustly forgotten. Later on Fay and Layzer [10], Kagan and Chubukov [14] generalized the Kohn-Lutinger results on the case of 3D Fermi gas with hard-core repulsion and got reasonable $T_C$'s of triplet p-wave pairing [11-14]

$$T_{C1} \sim \varepsilon_F \exp\left(-\frac{13}{f_0^2}\right), \qquad (1)$$

where $f_0 = 2p_F r_0/\pi$ is an effective gas parameter of Galitskii [75]. It should be noted that in their theory the Kohn's anomaly played important but not decisive role. According to the considerations of Prof. Nozieres [12], to get the attraction between two fermions in substance in the p-wave channel it is sufficient to have an effective potential which is increasing in the important interval from 0 to $2p_F$ in momentum space (and of course decreasing at larger momenta $p \geq 1/r_0$, where $r_0$ is the range of the potential). The authors got $T_C$ of the order of 1 mK for superfluid He-3 and (1–10) mK for electron plasma in metals in the limit of intermediate electron densities [15]. In the 2D case the authors predicted p-wave superconductivity below the critical temperature [11-13, 76-78]

$$T_{C1} \sim \varepsilon_F \exp\left(-\frac{1}{6.1 f_0^3}\right), \quad (2)$$

where $f_0 = 1/2 \ln(1/p_F r_0)$ is an effective 2D gas parameter of Bloom [79]. Note that specific form of polarization operator (and one-sided character of the Kohn's anomaly for $q \leq 2p_F$) $\text{Re}\sqrt{q-2p_F} = 0$ in 2D case prohibits the superconductive pairing in the second order of the gas parameter for the effective interaction $U_{\text{eff}}$. However in the third order of the gas parameter the character of the Kohn's anomaly changes on the opposite one $\text{Re}\sqrt{2p_F - q}$, and as a result a large 2D Kohn's anomaly becomes effective for superconductivity problem bringing nonzero critical temperature in Eq.2 [76]. A thorough evaluation of $T_C$ in 2D Fermi-gases (which includes the consideration of all irreducible diagrams of the third order in the gas parameter in the effective interaction) was fulfilled in [77, 78].

III) Superconductivity in repulsive-$U$ Hubbard model at low electron densities

Soon after the first papers on p-wave pairing, Baranov, Kagan and Chubukov [13, 80, 81] generalized these results on the lattice models. The authors considered 3D and 2D Hubbard model at low electron densities. In agreement with T-matrix Kanamori ideas [82] they came to the conclusion that Hubbard model at low density is equivalent to the Fermi-gas model with the effective gas parameters given by $f_0 = 2 p_F d / \pi$ ($d$ is intersite distance) in 3D and $f_0 = 1/2 \ln(1/p_F d)$ in 2D [83]. As a result they got the same $T_C$'s of triplet p-wave pairing as in the case of Fermi-gas under the substitution of $r_0$ on $d$ [82] in gas parameters in the Eqs. 1 and 2.

IV) Superconductivity in the Shubin-Vonsovsky model.

The Shubin-Vonsovsky model [7, 84-86] (or extended Hubbard model [84] as it is usually referred to in Western literature) is the most repulsive and thus the most unbeneficial model with respect to superconductivity. However even in this model in the strong coupling case $U \gg V \gg W$ (when both onsite Hubbard repulsion $U$ and intersite Coulomb repulsion $V$ are larger than the bandwidth $W$) a detailed analysis of [84] showed the appearance of triplet p-wave superconductivity in the 3D and 2D low density case. Moreover the main exponent for the Kohn-Luttinger critical temperature is given precisely by the same expression as for repulsive-$U$ Hubbard model in the limit $V = 0$. The presence of sufficiently large $V$ changes only the preexponential factor. In the opposite case of weak-coupling Born approximation $W \gg U \gg V$ the authors of [85-90] constructed the superconducting phase-diagrams which contain the regions of p-, d-, f- and anomalous s-wave pairing. The calculations were presented for quadratic and hexagonal lattices and can be important for superconductive pairing in FeAs-based superconductors [85-88] as well as for idealized monolayer and bilayer graphene. The crystal structure and typical phase diagram of the superconducting state in bilayer graphene [89,90] are presented on Figs. 1 and 2. Note that if we neglect the influence of the substrate potential and disregard structural disorder and both magnetic and nonmagnetic impurities, we get rather optimistic estimates for $T_C$'s in monolayer and especially bilayer graphene on the level of 10-20 K. [88-92]

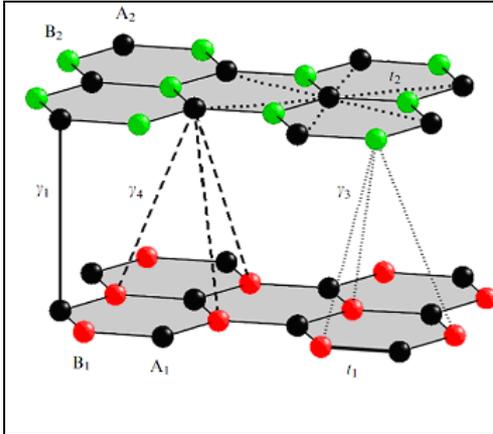 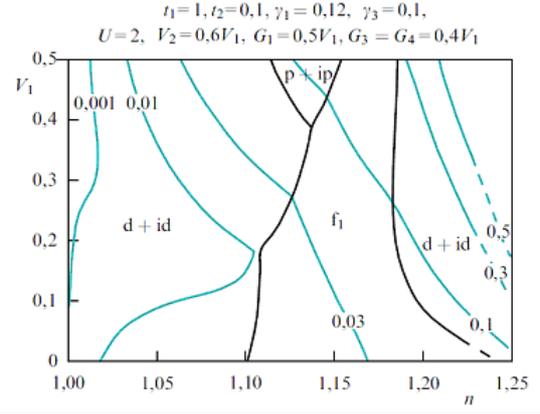

| Fig.1 Crystal structure of graphene bilayer [11, 89] | Fig.2 Phase diagram of superconducting state in idealized bilayer graphene [11, 89] |

V) Strong $T_C$ enhancement in the repulsive two-band Hubbard model and in a spin-polarized Fermi-gas

Usually the Kohn-Luttinger superconducting temperatures are low enough. There are two possible ways to increase $T_C$ considerably. First way is obvious. It is just to increase the density of electrons. The second possibility was proposed in [16-19, 93-97]. Namely the authors proposed to consider strongly spin-polarised Fermi gas [16, 19, 93] or the two-band situation [17, 94-97]. The strong $T_C$ increase in both cases already at low density is connected with the idea of the separation of the channels. According to this scenario the Cooper pair is formed by two fermions with spins up (by two electrons of the first band) while effective interaction for them is prepared by two fermions with spins down (by two electrons of the second band). In this case the Kohn's anomaly acquires a form $(q_\uparrow - 2p_{F\downarrow}) \ln|q_\uparrow - 2p_{F\downarrow}|$, where $q_\uparrow^2 = 2p_{F\uparrow}^2(1-\cos\theta)$.

In terms of the angle $\theta$ between incoming and outgoing momenta in the Cooper channel the Kohn's anomaly changes its form from $(\pi-\theta)^2 \ln(\pi-\theta)$ in the absence of spin polarization on $(\theta-\theta_C) \ln|\theta-\theta_C|$ in the presence of it, where $\theta_C$ differs from $\pi$ proportionally to $p_{F\uparrow}/p_{F\downarrow}-1$. Thus it becomes stronger when the ratio of the Fermi momentum of the two components $p_{F\uparrow}/p_{F\downarrow} \geq 1$ increases. In the same time the number (and the density of states) of down spins decreases with an increase of the spin-polarisation. As a result we have strongly nonmonotonous dependence of the triplet p-wave $T_C$ from the ratio of $p_{F\uparrow}/p_{F\downarrow} \geq 1$ or from the density ratio $n_\uparrow/n_\downarrow$. Moreover both in 3D and especially in 2D case (where we get now superconductivity already in the second order of the gas parameter for the effective interaction) we have a very pronounced and a very broad maximum for the optimal density ratio. In the 2D case the optimal density ratio $n_\uparrow/n_\downarrow = 4$ and the main exponent for triplet p-wave $T_C$ in maximum increases in several times bringing quite substantial values

$$T_{C1}^{\uparrow\uparrow} \sim \varepsilon_{F\uparrow} \exp\left(-\frac{1}{2f_0^2}\right). \qquad (3)$$

The dependence of $T_{C1}^{\uparrow\uparrow}$ in 2D polarised Fermi-gas with repulsion on the polarization degree $\alpha = (n_\uparrow - n_\downarrow)/(n_\uparrow + n_\downarrow)$ is presented on Fig.3.

This theory is important for spin-polarised He-3–He-4 mixtures [93] as well as for 3D and 2D imbalanced Fermi gases of $^6Li$ and $^{40}K$ in the reduced geometry of magnetic traps [19, 98]. For pure superfluid He-3 it predicts 6.4 times increase of triplet p-wave $T_C$ for superfluid A1-phase at optimal spin-polarisation of 48% [13, 16, 19, 93]. The 20% increase of the critical temperature in the A1-phase was obtained in magnetic fields $B = 15$ T in the experiments [99, 100].

Concerning electron systems, the theory predicts p-wave superconductivity in superlattices PbTe-SnTe and dichalcogenides $CuS_2$-$CuSe_2$ for geometrically separated bands belonging to the neighbouring layers [17].

It also predicts $T_C$ of the order of 0.5 K in very clean low density 2D heterostructures in a strong (parallel to the electronic layer) magnetic field [18]. The superconducting phase-diagram of the electronic monolayer in parallel magnetic field is presented on Fig.4.

Finally the theory [95-97] for superconductivity and electron polaron effect in the two band Hubbard model with one narrow band predicts strong enhancement of an effective mass of heavy electrons dressed in a virtual cloud of soft electron-hole pairs of light electrons. Moreover it yields quite reasonable $T_C$'s of the order of 5–10 K for pairing of heavy electrons in the situation when effective interaction for them is prepared by light electrons. The theory can be important for uranium based superconductors such as $UNi_2Al_3$ and some other mixed valence systems at low electron densities.

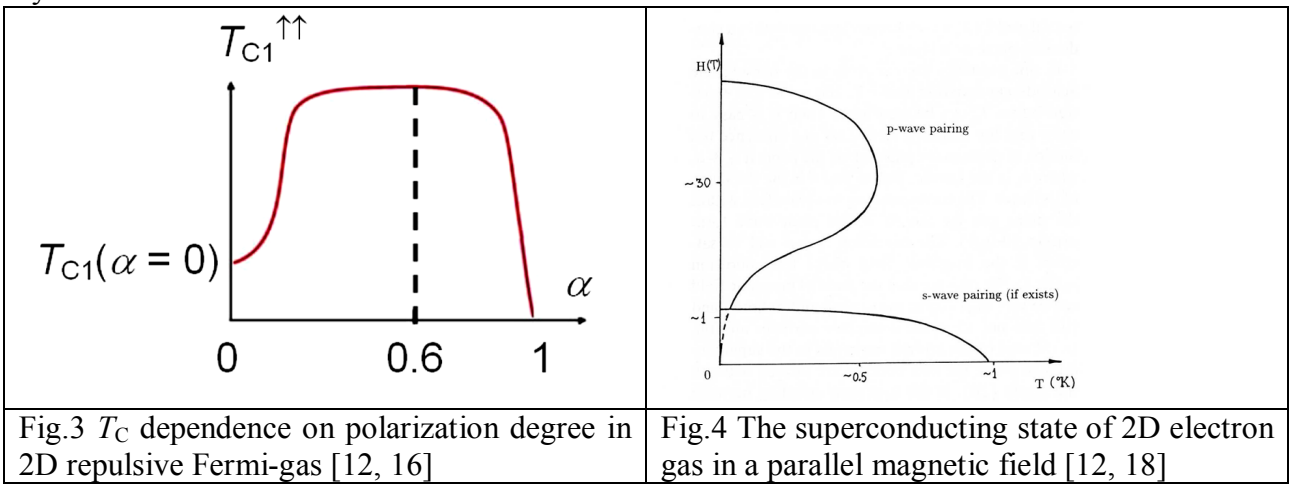

| Fig.3 $T_C$ dependence on polarization degree in 2D repulsive Fermi-gas [12, 16] | Fig.4 The superconducting state of 2D electron gas in a parallel magnetic field [12, 18] |

VI) High-$T_C$ superconductivity in the two-dimensional t-J model.

Usually Kohn-Luttinger mechanism and its generalizations are very effective to provide us with unconventional low temperature superconductivity in purely repulsive fermionic systems. To get really high $T_C$ superconductivity we should consider the models with not purely repulsive potentials. In this context the very important role belongs to the famous t-J model, which in fact is the model with the Van der Waals potential. It corresponds to strong onsite Hubbard repulsion and weak intersite AFM attraction. As it was shown by Kagan, Rice [24], it is very simple to get 100 K in this model for the set of parameters typical for high $T_C$ cuprates, namely for optimal electron densities $n_e \sim 0.85$ (optimal doping) and the ratios $J/t \sim 1/2$

$$T_C \sim \varepsilon_F \exp\left(-\frac{\pi t}{2 J n_e^2}\right). \qquad (4)$$

The symmetry of the order parameter in this case corresponds to $d_{x^2-y^2}$-type in agreement with experiments in cuprates. The interacting potential and superconducting phase diagram of the 2D t-J model iare presented on Figs. 5 and 6.

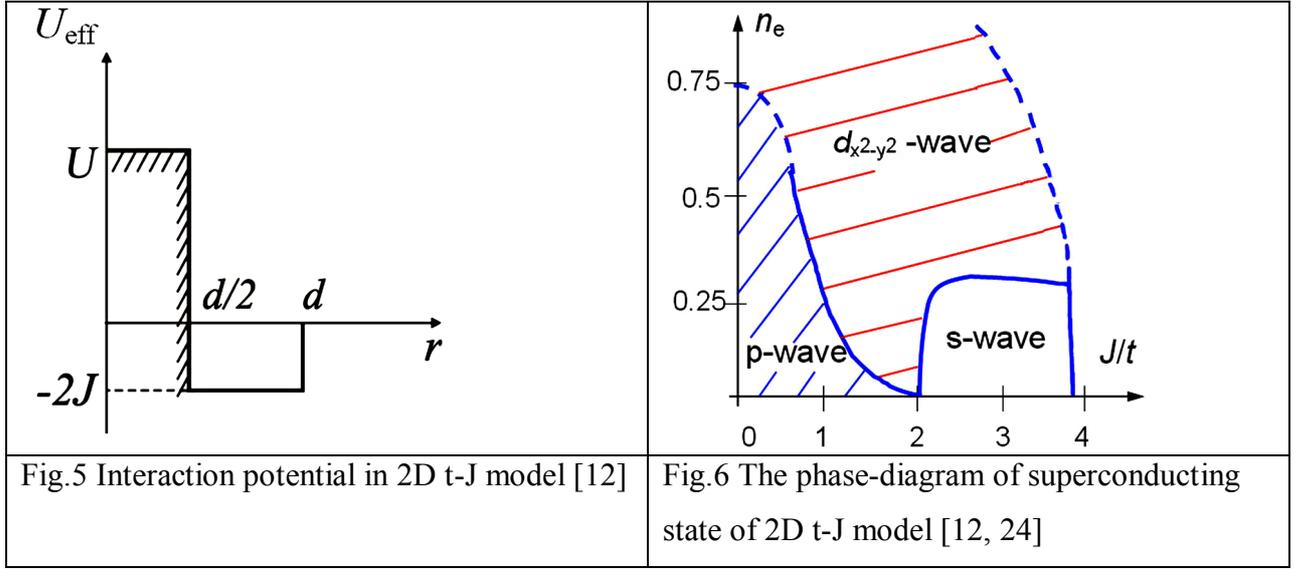

| Fig.5 Interaction potential in 2D t-J model [12] | Fig.6 The phase-diagram of superconducting state of 2D t-J model [12, 24] |

VII) Superconductivity and BCS-BEC crossover in attractive Fermi-gas and Hubbard models.

Let us go further on and consider purely attractive short-range potentials. In the absence of the lattice we have then an attractive Fermi-gas. In the lattice case we are dealing with the attractive-$U$ Hubbard model. Note that the attractive Hubbard model is a static model which does not contain the retardation effects. The phase-diagram of this model (as well as of the attractive Fermi-gas model) in the 3D case corresponds to the BCS-BEC crossover between extended Cooper pairs (formed in momentum space in the presence of filled Fermi-surface at small values of $|U|/W$) and local pairs or dimers (formed by two fermions in vacuum at large values of $|U|/W \geq 1$). In the BCS domain of extended pairs in both models the s-wave scattering amplitude corresponds to attraction $a < 0$, and in the same time the one-particle chemical potential is positive $\mu > 0$. Vise versa in the BEC domain of local pairs the scattering length is positive $a > 0$ (signaling the appearance of the two-particle bound state in vacuum) and correspondingly the chemical potential is negative ($\mu < 0$). Moreover for the BEC domain we have two characteristic temperatures. The higher crossover (Saha) temperature [30] is given by:

$$T^* \sim \frac{|E_B|}{3/2 \ln\left(\frac{|E_B|}{\varepsilon_F}\right)}, \qquad (5)$$

where $|E_B|$ is a binding energy of a local pair. Let us emphasize that $T^*$ in Eq. 5 corresponds to the thermodynamic equilibrium between creation and dissociation of local pairs [101]. In the same time the lower (BEC) critical temperature corresponds to the real phase-transition and in the principal approximation is given by Einstein formula [102]:

$$T_C^{BEC} \sim 3.31 \left(\frac{n}{2}\right)^{2/3} / m_B, \qquad (6)$$

where $m_B$ is a bosonic mass (a mass of a local pair or a dimer). In the strong-coupling limit $|U|>W$ in a lattice model with discrete hoppings only on neighbouring sites a bosonic mass will be enhanced $m_B \sim m \frac{|U|}{W}$ according to the second order perturbation theory of [27] (at first hops one electron of the local pair, virtually destroying it, and after that the second one restoring the pair). Note that in the absence of the lattice (in Fermi-gas model) $m_B = 2m$. For the case of Fermi-gas substituting in Eq.6 the 3D fermionic density $n = p_F^3/3\pi^2$ and Fermi-energy $\varepsilon_F = p_F^2/2m$ we finally get:

$$T_C^{BEC} \sim 0.2\varepsilon_F[1+1.3a_{2-2}n^{1/3}], \qquad (7)$$

where nontrivial corrections to Einstein expression [103] are governed by the dimer-dimer scattering amplitude. In exact calculations of Petrov et al., and Brodsky et al., $a_{2-2} = 0.6a$ [104-107]. As we already mentioned in the Introduction, the BCS-BEC crossover experimentally was realized in ultracold Fermi-gases in the regime of Feshbach resonance [31-39]. The typical phase-diagram of the BCS-BEC crossover in 3D resonance Fermi-gas is presented on Fig. 7.

Note that in between the two characteristic temperatures, namely for $T_C \ll T \ll T^*$, we have an interesting new phase of a normal Bose metal. Note also that an important criterion for the stability of the local pairs state is connected with the low density limit, or in another words, with the condition $|E_B| > \varepsilon_F$. This condition means that the effective radius of the local pair $a$ $\left(|E_B| = 1/ma^2\right)$ is much smaller than the interparticle distance $1/p_F$.

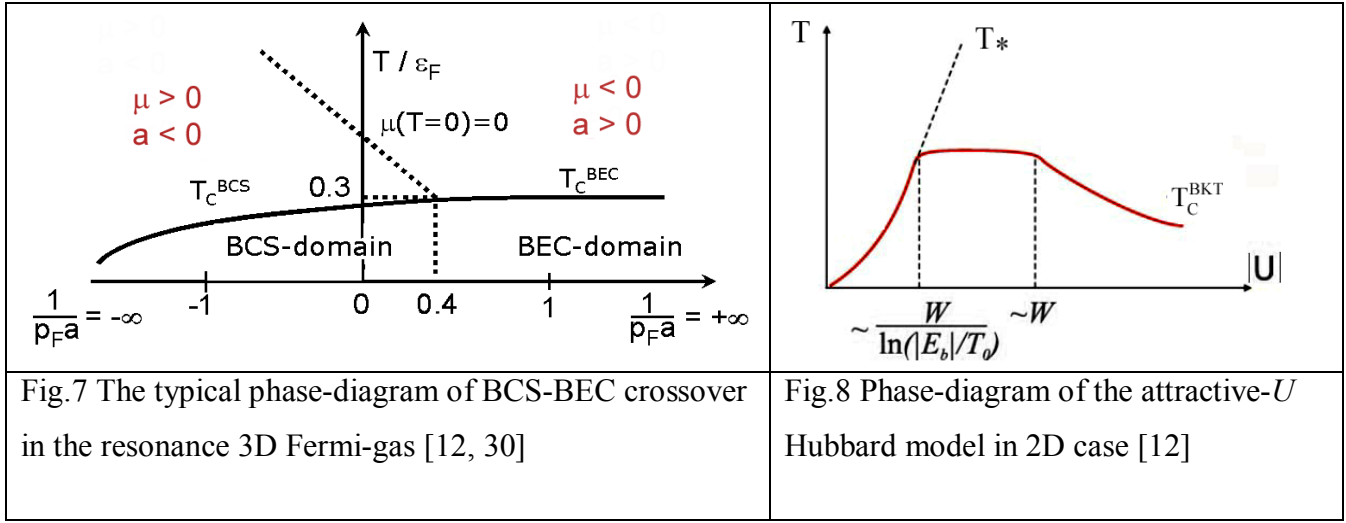

| Fig.7 The typical phase-diagram of BCS-BEC crossover in the resonance 3D Fermi-gas [12, 30] | Fig.8 Phase-diagram of the attractive-$U$ Hubbard model in 2D case [12] |

When we increase the density, the local pairs at first (for $|E_B| \sim \varepsilon_F$) begin to touch each other and finally (at large densities) crush the paired state completely. At intermediate densities for $1 \leq ap_F \leq 3$ in BEC domain [12] we can describe the system in terms of Fermi-Bose mixture of local pairs and one particle excitations. Note that in dilute BCS limit the critical temperature is given by Gor'kov-Melik-Barkhudarov formula for 3D attractive Fermi-gas [108]:

$$T_C = 0.28\varepsilon_F \exp\left(-\frac{1}{|f_0|}\right), \qquad (8)$$

where a 3D gas parameter $|f_0| = 2|a|p_F/\pi$.

The situation in 2D attractive-$U$ Hubbard model as well as in the 2D Fermi-gas model is qualitatively similar to the 3D case in a low density limit. The only substantial difference is connected with an appearence of the intermediate coupling case which is realized in the parameter region $\frac{W}{\ln(|E_B|/\varepsilon_F)} < |U| < W$ (see [12] and Fig. 8). In 2D Fermi-gas model according to the Fisher-Hohenberg theory [109] the mean field (BEC) critical temperature in the local pairs domain is given by

$$T_C^{BEC} \sim \frac{\varepsilon_F}{4\ln\left(1/f_{2-2}\right)}, \qquad (9)$$

where according to [110] $f_{2-2} \sim 1/\ln(1.6|E_B|/\varepsilon_F)$ is a 2D coupling constant which describes the repulsive interaction between the dimers. In low density 2D Fermi-gas the mean field $T_C^{BEC}$ in

Eq. 9 is only slightly different from the exact Berezinskii-Kosterlitz-Thouless critical temperature [111, 112]:

$$\frac{|T_C^{BEC} - T_C^{BKT}|}{T_C^{BEC}} \sim f_{2-2} \ll 1. \quad (10)$$

The crossover Saha temperature in 2D yields:

$$T^* \sim \frac{|E_B|}{\ln\left(\frac{|E_B|}{\varepsilon_F}\right)}, \quad (11)$$

where a binding energy is given by $|E_B| = \dfrac{1/md^2}{\exp\left(\dfrac{4\pi}{md^2|U|}\right) - 1} \sim \dfrac{W}{\exp\left(\dfrac{W}{|U|}\right) - 1}$.

For intermediate temperatures $T_C^{BKT} \ll T \ll T^*$ we again have a phase of normal bosonic metal. In 2D attractive-$U$ Hubbard model in this phase even in a very clean limit (no impurities) the resistivity behaves as $R(T) \sim \sqrt{T}$ (due to boson-boson scattering with Umklapp processes on the lattice [12, 25, 26]). Note that resistivity characteristics of this type can be obtained in degenerate semiconductors. Also quite interesting is a tunneling process between normal bosonic metal and standard BCS superconductor [12].

In the same time the BCS critical temperature in 2D is given by Miyake formula [113]

$$T_C^{BCS} \sim \sqrt{2|E_B|\varepsilon_F}. \quad (12)$$

Note that for symmetric potentials in 2D the two-particle bound states appear already at infinitely small attraction [114] (in contrast with the 3D case where we have a threshold for bound state formation [115]). So, the two phenomena coexist: the Cooper pairing in momentum space (in the presence of filled Fermi-sphere) and the formation of the local pair in real space in vacuum [116, 117]. Let us stress that in the BCS domain creation and Bose condensation of Cooper pairs take place simultaneously, and thus $T_C^{BCS} = T^*$.

Note also that for the weak-coupling case $|E_B| \ll \varepsilon_F$, as it was shown in [118], the mean-field $T_C^{BCS}$ in Eq. (12) is close again to exact critical temperature $T_C^{BKT}$:

$$\frac{|T_C^{BCS} - T_C^{BKT}|}{T_C^{BCS}} \sim \frac{T_C^{BCS}}{\varepsilon_F} \sim \sqrt{\frac{|E_B|}{\varepsilon_F}} \ll 1. \quad (13)$$

For the intermediate coupling case $|E_B| \leq \varepsilon_F$ and difference between $T_C^{BCS}$ and $T_C^{BKT}$ becomes more substantial. Let us emphasise that in the BCS domain at $T = 0$ we have a positive chemical potential $\mu = \varepsilon_F - |E_B|/2 > 0$ [113].

VIII) Strong coupling Migdal-Eliashberg theory

The recordly high conventional superconductivity in metallic H$_2$S under pressure brings our attention back to more standard electron-phonon mechanisms. Note that the classical strong coupling Migdal-Eliashberg theory [51, 52] is based on the concept of adiabaticity and assumes that both the coupling constant of electron-phonon interaction is not very large ($\lambda \leq 1$) and the ratio between Debye frequency and Fermi-energy is small $\omega_D/\varepsilon_F < 1$. The theory works pretty well in many conventional superconductors discovered in the past as well as in several new systems like K$_3$C$_{60}$ [69-71], metallic hydrides [1-5] and possibly in metallic hydrogen, which is not discovered yet.Note that in 1968 the very high $T_C$ superconductivity of electron-phonon nature was proposed by Aschcroft [61,62] for metallic hydrogen at high pressures.

IX) Alexandrov-Ranninger bipolaron theory in the nonadiabatic systems with strong polaron effects

Challenging Migdal-Eliashberg theory [51, 52], Alexandrov, Ranninger in 1981 [119, 120] advanced the bipolaron theory based on nonadiabaticity and strong polaronic effects. Note that stability of Alexandrov-Ranninger bipolaron state in total analogy with attractive-$U$ Hubbard model requires a low density limit for charge carriers (polarons and bipolarons). In another words, a bipolaron binding energy $|E_B|$ should be larger than the effective width of the narrow band $W_{eff}$ (strongly reduced by polaron effect): $|E_B| > W_{eff}$ [121]. Moreover the material should be characterized by large electron-phonon coupling constant $\lambda > 1$.

Experimentally the bipolaron theory should correspond to large effective masses of small radius bipolarons and the systems with an unconventional negative curvature (with a strong increase) of the upper critical field $H_{C2}$ at low temperatures $T \ll T_C$ [122, 123]. Note that in layered high $T_C$ cuprates the thorough investigations of the phonon spectrum by Andersen group [124] yield the electron-phonon coupling constant $\lambda \sim 1/2$ even for the bismuth family of cuprates (where strong-coupling effects were mostly probable), so bipolaron formation is unlikely here.

X) Superconductivity in the Fermi-Bose mixture model

The conditions for local pairs formation and their stability (as we already mention in section VII) are milder in the model of the two component Fermi-Bose mixture firstly introduced by Ranninger et al. [40, 41]. This model was initially formulated for cuprates (see also an interesting suggestion of Geshkenbein, Ioffe, Larkin [42] who advocated this model for 2D electron systems with filling factors close to Van Howe singularities on quadratic lattice or for quasi-1D spin-ladder systems). However later on it found an important application for the ultracold quantum gases in the regime of Feshbach resonance. In the last case it is usually called the two channel Feshbach model [39] and contains a Feshbach-Ranninger term which effectively transforms the Cooper pair of two fermions in one channel into the real boson (a local pair) in the other channel [39, 125, 126].

Note that the formation of local pairs of the excitonic nature is possibly realized in $BiO_6$ clusters in superconducting alloys $BaKBiO_3$ due to valence skipping considerations of Varma [126]. Moreover according to Kagan, Menushenkov et al., [44, 45]. In this class of materials we can possibly describe the system in the framework of the two component Fermi-Bose mixture of Alexandrov, Ranninger type, but with one substantial remark that the two components are separated in the real space, while the separation between them in energy space is absent. This model describes rather well many transport and thermodynamic properties of these materials. The crystalline structure of $BaKBiO_3$ and the interplay between Fermi and Bose subsystems are presented on Figs 9 and 10. Note that superconductivity with a critical temperature $T_C \sim 36$ K in this class of materials is connected with the delocalization and phase-coherence in the bosonic subsystem which is achieved via sucsssesive processes of coherent tunneling of local pairs from one Bose cluster to a neighbouring one (and so on) through the effective barriers prepared by Fermi clusters.

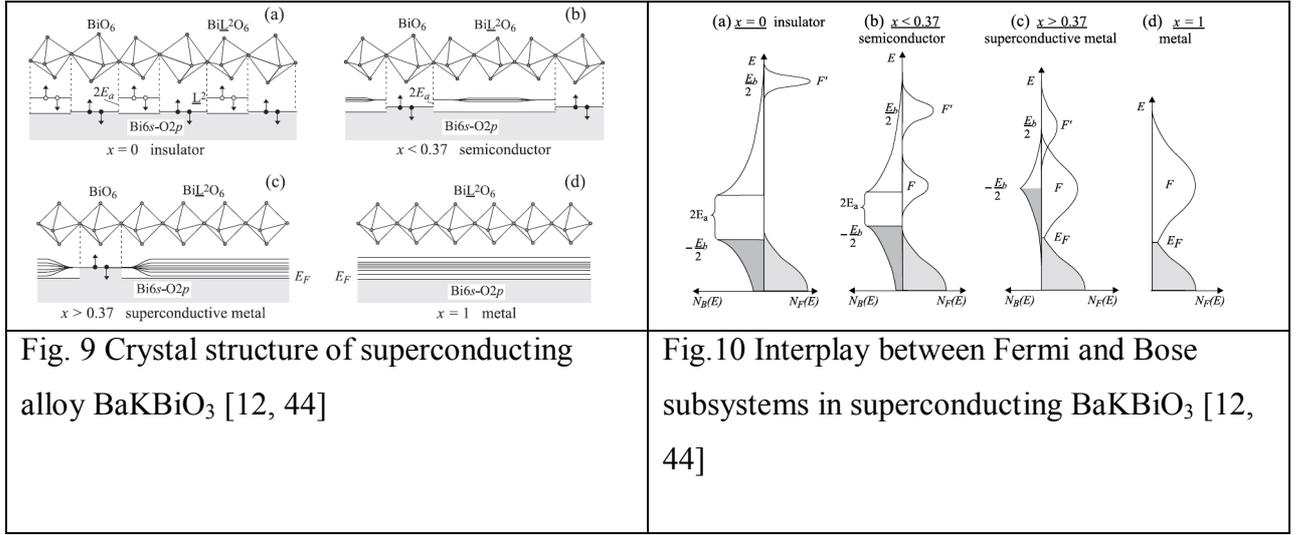

Fig. 9 Crystal structure of superconducting alloy BaKBiO$_3$ [12, 44]

Fig.10 Interplay between Fermi and Bose subsystems in superconducting BaKBiO$_3$ [12, 44]

XI) Discussion. Hidden unconventional superconductivity in graphene and conventional superconductivity in metallic hydrogen alloys under pressure

As we already mentioned in section IV, the Kohn-Luttinger mechanism gives rather optimistic predictions for idealized monolayer and bilayer graphene. In contrast to these expectations in real graphene superconductivity has not been observed yet experimentally. The reason for that is probably connected with the presence of structural disorder and nonmagnetic impurities in the real graphene. Even record experimental parameters with respect to the maximal effective mean free path $l \sim 2000$ Å and the maximal 2D electron density $n \sim 10^{13} cm^{-2}$ correspond today to the so-called dirty limit (or in the best case to the moderately clean limit), where the anomalous Kohn-Luttinger superconductivity will be totally suppressed by disorder in agreement with Abrikosov-Gor'kov-Larkin theory [125, 128, 129] (see also [130-135] for the new results on the interplay between superconductivity and localization). Thus the experimental challenge for the future investigations of graphene is to go to the superclean limit, which means either to reduce the degree of disorder and the number of impurities or to increase the electron density in the monolayer. One more option here can be connected with quasi one-dimensional structures such as graphene nanoribbons. In this structures, however, we should expect the competition between superconducting ordering and Peierls-like instabilities of SDW or CDW type [136]. Finally Volovik and Esquinazi [137] advocated the idea of topologically protected superconductivity on the special graphite interfaces.

Note that the role of disorder is very interesting also in the attractive-$U$ Hubbard model. Here according to BdG analysis [47, 48, 138] of Randeria et al., [135] at strong disorder and low temperatures $T \to 0$ we will get the superconductive islands separated by an insulating sea. Moreover in the strong-coupling limit of the attractive-$U$ Hubbard model (for $|U|/W \gg 1$) the insulating phase corresponds to the Bose glass phase in accordance with the prediction of Fisher [139]. This result is very important for superconductor-insulator transition in disordered thin films [140, 141].

The conventional mechanisms can be also rather effective especially in the systems with high phonon frequencies ω and relatively large coupling constant λ ~ 1 (which is not compensated in these systems by the small Coulomb pseudopotential μ*), as well as in the systems with anharmonic two-well potentials [142]. The metallic hydrogen stabilized by sulfur in H$_3$S at high pressures and possibly not experimentally discovered yet pure metallic hydrogen could be the very good candidatures for the Migdal-Eliashberg mechanism of superconductivity and its generalizations.

Note that historically the first discussion on the phase transition from molecular to metallic hydrogen phase at high pressures belong to Wigner and Huntington in 1935 [57]. We should also mention here important papers by Kronig, de Boer and Korringa [58] and by Abrikosov [59, 60].

Thorough examination of the important families of the local minima of the thermodynamic potential for metallic hydrogen at different pressures in a normal (non-superconducting) state belongs to Brovman, Yu. Kagan and Kholas [63, 64]. At relatively low pressures and $T = 0$ they predict strongly anisotropic triangular phase with proton filaments in an electron liquid. These filaments form rigid 2D triangular lattice in the plane perpendicular to them. In the same time (similar to vortex lattice in type-II superconductor or in superfluid He-4) there is a tendency towards liquid-like motion in the direction parallel to the filaments. At higher pressures the local minima correspond in their calculations to planar structures. These structures also have pronounced liquid-like tendencies which (similar to graphite) are connected with the smallness of the effective shear modulus. Finally at very high pressures the local minima correspond to the structures which become more and more isotropic when we increase the pressure.

In this context an important unresolved problem is to construct the gross phase diagram of solid hydrogen under pressure and to determine the global minima both in molecular and metallic phases as well as the position of the transition line $P^*(T)$ between them. The main difficulty here is connected with a big uncertainty in the calculation of the thermodynamic potential for molecular phase. Here we should mention the first experimental results of Eremets, Troyan, Drozdov [2]. They get the possible transition from molecular to the metallic phase at the pressures of the order $P \sim 360$ GPa and temperatures 200 K.

XII) Conclusions.

During last 30 years after discovery of high-$T_C$ superconductivity we observe a lively discussion between different theoretical schools which support conventional or unconventional scenarious of superconductivity. The experimentalists are constantly supplying the both parties with the new evidences and new superconducting families which belong either to conventional or unconventional class of materials. The dispute clearly enriches the fascinating and rapidly developing field of high-$T_C$ superconductivity. The broad and objective (unbiased) outlook is necessary for the new predictions on the most perspective materials with critical temperatures approaching the room temperature superconductivity.

XIII) The author acknowledges stimulating discussions with many colleagues and especially grateful to N.M. Plakida, G.V. Shlyapnikov, A.S. Alexandrov*, A.V. Chubukov, M.A. Baranov, D.V. Efremov, M.V. Feigel'man, I.S. Burmistrov, V.V. Val'kov, V.A. Mitskan, M.M. Korovushkin, K.I. Kugel, A.P. Menushenkov and A.Ya.Tzalenchuk. M.Yu.K acknowledges support in the framework of the RFBR grant N 1402.0058 and thanks the Program of Basic Research of the National Research University Higher School of Economics.

References

[1]. A.P. Drozdov, M.I. Eremets, I.A. Troyan, Nature, **525**, 73 (2015)
[2]. M.I. Eremets, I.A. Troyan, A.P. Drozdov, cond-mat 1601.04479 (2016)
[3]. Y. Wang, Y. Ma, J. Chem. Phys., **140**, 040901 (2014)
[4]. D. Duan, Y. Liu, F. Tian, D. Li, X. Huang, Z. Zhao, H. Yu, B. Liu, W. Tian, T. Cui, Scient. Reports, **4**, 6968 (2014)
[5]. Y. Liu, D. Duan, F. Tian, D. Li, X. Sha, Z. Zhao, H. Zhang, G. Wu, H. Yu, B. Liu, T. Cui, cond-mat 1503.08587 (2015)
[6]. J.C. Hubbard, Proc.Royal Soc.Lond., **A276**, 238 (1963)
[7]. S.P. Shubin, S.V. Vonsovsky, Phys. Z. Sowjetunion, **7**, 292 (1935)
[8] S.P. Shubin, S.V. Vonsovsky, Phys. Z. Sowjetunion, **10**, 348, (1936)

[9]. W. Kohn, J.M. Luttinger, Phys.Rev.Lett., **15**, 524 (1965)
[10]. D. Fay, A. Layzer, Phys. Rev. Lett., **20**, 187 (1968)
[11]. M.Yu. Kagan, M.M. Korovushkin, V.A. Mitskan, Physics Uspekhi, **58**, 733 (2015)
[12]. M.Yu. Kagan, Modern Trends in Superconductivity and Superfluidity, Springer, Dordrecht, 2013, Lecture notes in Physics, v.874.
[13]. M.A. Baranov, A.V. Chubukov, M.Yu. Kagan, Int.Jour. Mod. Phys. B, **6**, 2471 (1992)
[14]. M.Yu. Kagan, A.V. Chubukov, JETP Lett., **47**, 525 (1988)
[15]. A.V. Chubukov, M.Yu. Kagan, Jour.Phys.: Condens.Matt., **1**, 3135 (1989)
[16]. M.Yu. Kagan, A.V. Chubukov, JETP Lett., **50**, 483, (1989)
[17]. M.Yu. Kagan, Phys. Lett.A, **152**, 303, (1991)
[18]. M.A. Baranov, D.V. Efremov, M.Yu. Kagan, Physica C: Superconductivity, **218**, 75 (1993)
[19]. M.A. Baranov, M.Yu. Kagan, Yu. Kagan, JETP Lett., **64**, 301 (1996)
[20]. I.E. Dzyaloshinskii, V.M. Yakovenko, Sov. Phys. JETP, **94**, 344 (1988)
[21]. A.I. Kozlov, Supercond.: Phys. Chem. Eng., **2**, 64 (1989)
[22]. K.I. Hur, T.M. Rice, Ann.Phys., **325**, 1452 (2009)
[23]. V.J. Emery, S.A. Kivelson, H.Q. Lin, Phys. Rev. Lett., **64**, 475 (1990)
[24]. M.Yu. Kagan, T.M. Rice, Jour. Phys.: Condens.Matt., **6**, 3771 (1994)
[25]. M.Yu. Kagan, R. Fresard, M. Capzzali, H. Beck, Phys.Rev.B, **57**, 5995 (1998)
[26]. M.Yu. Kagan, R. Fresard, M. Capzzali, H. Beck, Physica B, **284-288**, 347 (2000)
[27]. P. Nozieres, S. Schmitt-Rink, Jour. Low Temp. Phys., **59**, 195 (1985)
[28]. A.J. Leggett, J. Phys. (Paris) Colloq., **41**, C7 (1980)
[29]. A.J. Leggett, in Modern trends in the Theory of Condensed Matter., eds.A. Pekalski, J. Przystawa, Springer, Berlin, 1980, Lecture notes of the XVI Karpacz Winter school of Theoretical physics
[30]. R. Combescot, X. Leyronas, M.Yu. Kagan, Phys.Rev. A, **73**, 023618 (2006)
[31]. H. Feshbach, Ann.Phys., **5**, 357 (1958)
[32]. H. Feshbach, Ann.Phys., **19**, 287 (1962)
[33]. U. Fano, Nuovo Cimento, **12**, 156 (1935)
[34]. U. Fano, Phys.Rev., **124**, 1866 (1961)
[35]. C.A. Regal, C. Ticknor, J. L. Bohn, D. S. Jin, Nature, **424**, 47 (2003)
[36]. C.A. Regal, M. Greiner, D. S. Jin, Phys. Rev. Lett., **92**, 040403 (2004)
[37]. M.W. Zwierlein, C.A. Stan, C.H. Schunck, S.M.F. Raupach, S. Gupta, Z. Hadzibabic, W. Ketterle,, Phys. Rev. Lett., **91**, 250401 (2003)
[38]. M.W. Zwierlein C.A. Stan, C.H. Schunck, S.M.F. Raupach, A.J. Kerman, W. Ketterle, Phys. Rev. Lett., **92**, 120403 (2004)
[39]. V. Gurarie, L. Radzihovsky, Ann.Phys.(Weinheim), **322**, 2 (2007)
[40]. J. Ranninger, R. Micnas, S. Robaszkiewicz, Ann.Phys., **13**, 455 (1988)
[41]. R. Micnas, J. Ranninger, S. Robaszkiewicz, Rev. Mod. Phys., **62**, 113, (1990)
[42]. V.B. Geshkenbein, L.B. Ioffe, A.I. Larkin, Phys. Rev. B, **55**, 3173, (1997)
[43]. M.Yu. Kagan, I.V.Brodsky, D.V. Efremov, A.V. Klaptsov, Phys. Rev. A, **70**, 023607 (2004)
[44]. M.Yu. Kagan, A.P. Menushenkov, A.V. Kuznetsov, K.V. Klement'ev, Sov. Phys. JETP, **93**, 615, (2001)
[45]. A.P. Menushenkov, A.V. Kuznetsov, K.V. Klement'ev, M.Yu. Kagan, Jour. Supercond. Nov. Magn., **29**, 701, (2016)
[46]. J. Barden, L.N. Cooper, J.R. Schrieffer, Phys. Rev., **108**, 1175 (1957)
[47]. N.N. Bogoliubov, Sov. Phys. JETP, **34**, 58 (1958)
[48]. N.N. Bogoliubov, Sov. Phys. JETP, **34**, 73 (1958)
[49]. L.P.Gor'kov, Developing BCS ideas in the former Soviet Union, In BCS: 50 years, eds. L.N. Cooper, D. Feldman, World Scientific, 2011
[50]. L.P.Gor'kov, Int. Jour. Mod. Phys. B, **24**, 3835 (2010)
[51]. A.B. Migdal, Sov. Phys. JETP, **34**, 958 (1958)

[52]. G.M. Eliashberg, Sov. Phys. JETP, **11**, 3 (1960)
[53]. W.L. Mcmilan, Phys. Rev. B, **167**, 331 (1968)
[54]. R.C. Dynes, Solid State Commun., **10**, 615 (1972)
[55]. P.B. Allen, R. C. Dynes, Phys. Rev. B, **12**, 905 (1975)
[56]. R. Combescot, Phys. Rev. Lett., **67**, 148 (1991)
[57]. E. Wigner, H.B. Huntington, J. Chem. Phys., **3**, 764 (1935)
[58]. R. Kronig, J.de Boer, J. Korringa, Physica (Utrecht), **12**, 245 (1946)
[59]. A.A. Abrikosov, Astron. Zh., **31**, 112 (1954)
[60]. A.A. Abrikosov, Sov. Phys. JETP, **12**, 1254 (1961)
[61]. N.W. Aschkroft, Phys. Rev. Lett., **21**, 1748 (1968)
[62]. N.W. Aschkroft, Phys. Rev. Lett., **92**, 187002 (2004)
[63]. E.G. Brovman, Yu. Kagan, A. Kholas, Sov. Phys. JETP, **34**, 1300 (1972)
[64]. E.G. Brovman, Yu. Kagan, A. Kholas, Sov. Phys. JETP, **35**, 783 (1972)
[65]. P. Cudazzo, G. Profeta, A. Sanna, A. Floris, A. Continenza, S. Massidda, E.K.U. Gross, Phys. Rev. B, **81**, 134506 (2010)
[66]. N. Bernstein, C. Hellberg, M. Johannes, I. Mazin, M.J. Mehl, Phys. Rev. B, **91**, 060511(R) (2015)
[67]. I. Errea, M. Calandra, C.J. Pickard, J. Nelson, R.J. Needs, Y. Li, H. Liu, Y. Zhang, Y. Ma, F. Mauri, Phys. Rev. Lett., **114**, 157004 (2015)
[68]. M.I. Eremets, I.A. Trojan, S.A. Medvedev, J.S. Tse, Y. Yao, Science, **319**, 1506 (2008)
[69]. A. Bianconi, T. Jarlborg, Europhys. Lett., **112**, 37001 (2015)
[70]. A. Bianconi, T. Jarlborg, Nov. Supercond. Mat., **1**, 37 (2015)
[71]. T. Jarlborg, A.Bianconi, Scient. Rep., **6**, 24816 (2016)
[72]. W. Kohn, Phys. Rev. Lett., **2**, 393 (1959)
[73]. J. Friedel, Adv. Phys., **3**, 446 (1954)
[74]. J. Friedel, Nuovo Cimento Suppl., **7**, 287 (1958)
[75]. V.M. Galitskii, Sov. Phys. JETP, **7**, 104 (1958)
[76]. A.V. Chubukov, Phys. Rev.B, **48**, 1097 (1993)
[77]. D.V. Efremov, M.S. Mar'enko, M.A. Baranov, M.Yu. Kagan, Physica B, **284-288**, 216, (2000)
[78]. M.A. Baranov, D.V. Efremov, M.S. Mar'enko, M.Yu. Kagan, Sov. Phys. JETP, **90**, 861 (2000)
[79]. P. Bloom, Phys. Rev.B, **12**, 125 (1975)
[80]. M.A. Baranov, M.Yu. Kagan, Zeit. Phys. B: Condens. Matt., **86**, 237 (1992)
[81]. M.A. Baranov, M.Yu. Kagan, Sov. Phys. JETP, **99**, 1236 (1991)
[82]. J. Kanamori, Progr. Theor. Phys., **30**, 275 (1963)
[83]. H. Fukuyama, Y. Hasegawa, O. Narikiyo, Jour. Phys. Soc. Jpn., **60**, 2013 (1991)
[84]. M.Yu. Kagan, D.V. Efremov, M.S. Mar'enko, V.V. Val'kov, JETP Lett., **93**, 819 (2011)
[85]. M.Yu. Kagan, V.V. Val'kov, M.M. Korovushkin, V.A. Mitskan, JETP Lett., **97**, 236 (2013)
[86]. M. Yu. Kagan, V.V. Val'kov, V.A. Mitskan, M.M. Korovushkin, Sov. Phys. JETP, **117**, 728 (2013)
[87]. M. Yu. Kagan, V.V. Val'kov, V.A. Mitskan, M.M. Korovushkin, Sov. Phys. JETP, **118**, 995 (2014)
[88]. M.Yu. Kagan, V.V. Val'kov, V.A. Mitskan, M.M. Korovushkin, Solid State Commun., **188**, 61 (2014)
[89]. M.Yu. Kagan, V.A. Mitskan, M.M. Korovushkin, Eur. Phys. Jour. B, **88**, 157 (2015)
[90]. M.Yu. Kagan, V.A. Mitskan, M.M. Korovushkin, ZHETP, **146**, 1301 (2014)
[91]. M.Yu. Kagan, V.A. Mitskan, M.M. Korovushkin, Jour. Supercond. Nov. Magn., **29**, 1043, (2016)
[92]. M.Yu. Kagan, V.A. Mitskan, M.M. Korovushkin, cond-mat 1511.00377 (2016)
[93]. M.Yu. Kagan, Physics Uspekhi, **37**, 69 (1994)

[94]. M.A. Baranov, M.Yu. Kagan, Sov. Phys. JETP, **75**, 165 (1992)
[95]. M.Yu. Kagan, V.V. Val'kov, Low Temp. Phys., **37**, 84 (2011)
[96]. M.Yu. Kagan, V.V. Val'kov, in "A Lifetime in Magnetism and Superconductivity: A Tribute to Professor David Shoenberg", Cambridge Scientific Publishers, 2011
[97]. M.Yu. Kagan, V.V. Val'kov, ZHETP, **140**, 196 (2011)
[98]. W. Ong, C.-Y. Cheng, I. Arakelyan, J.E. Thomas, Phys. Rev. Lett., **114**, 110403 (2015)
[99]. G. Frossati, K.S. Bedell, S.A.J. Wiegers, G.A. Vermeulen, Phys. Rev. Lett., **57**, 1032 (1986)
[100]. G. Frossati et al, Czech. J.Phys., v. 440, p.909, (1990)
[101]. L.D. Landau, E.M. Lifshitz, Statistical Physics, Part I, Butterworth-Heinemann, Oxford, 1980
[102]. A. Einstein, Sber. Preuss. Akad. Wiss., **1**, 3 (1925)
[103]. V.A. Kashurnikov, N.V. Prokof'ev, B.V.Svistunov, Phys.Rev. Lett., **87**, 120402 (2001)
[104]. D.S. Petrov, C. Salomon, G.V. Shlyapnikov, Phys. Rev. A, **71**, 012708 (2005)
[105]. I.V. Brodsky, M.Yu. Kagan, A.V. Klaptsov, R. Combescot, X. Leyronas, JETP Lett., **83**, 306 (2006)
[106]. I.V. Brodsky, M. Yu. Kagan, A.V. Klaptsov, R. Combescot, X. Leyronas, Phys. Rev. A, **73**, 032724 (2006)
[107]. M.Yu. Kagan, I.V. Brodsky, A.V. Klaptsov, R. Combescot, X. Leyronas, Physics Uspekhi, **176**, 1105 (2006)
[108]. L.P. Gor'kov, T.K. Melik-Barkhudarov, Sov. Phys. JETP, **40**, 1452 (1961)
[109]. D.S. Fisher, P.C. Hohenberg, Phys. Rev. B, **37**, 4936 (1988)
[110]. D.S. Petrov, M.A. Baranov, G.V. Shlyapnikov, Phys. Rev. A, **67**, 031601 (2003)
[111]. J.M. Kosterlitz, D.J. Thouless, Jour. Phys.C: Solid State Phys., **6**, 1181 (1973)
[112]. V.L. Berezinskii, JETP Lett., **34**, 610 (1972)
[113]. K. Miyake, Progr. Theor. Phys., **69**, 1794 (1983)
[114]. D.V.Efremov, M.Yu. Kagan, Physica B, **329-333**, 30 (2003)
[115]. L.D. Landau, E.M. Lifshitz, Quantum Mechanics: Non-Relyativistic Theory, Pergamon Press, UK, 1977
[116]. M. Randeria, J.M. Duan, L.Y. Shieh, Phys. Rev. Lett., **62**, 981 (1989)
[117]. S. Schmitt-Rink, C.M. Varma, A.E. Ruckenstein, Phys. Rev. Lett., **63**, 445 (1989)
[118] M.R. Beasley, J.E. Mooij, T.P. Orlando, Phys. Rev. Lett., **42**, 165 (1979)
[119]. A.S. Alexandrov, J. Ranninger, Phys. Rev. B, **23**, 1796 (1981)
[120]. A.S. Alexandrov, J. Ranninger, Phys. Rev. B, **24**, 1164 (1981)
[121]. A.S. Alexandrov, D.A. Samarchenko, S.V. Traven, Sov. Phys. JETP, **66**, 567 (1987)
[122]. A.S. Alexandrov, Phys. Rev. B, **38**, 925 (1988)
[123]. A.S. Alexandrov, Physica C: Superconductivity, **158**, 337 (1989)
[124]. O.K. Andersen, A.I. Liechtenstein, O. Rodriguez, I.I. Mazin, O. Jepsen, V.P. Antropov, O. Gunnarsson, S. Gopalan, Physica C: Superconductivity, **185-189**, 147 (1991)
[125]. E.A. Donley, N.R. Claussen, S.T. Thompson, C.E. Wieman, Nature, **417**, 529(2002)
[126]. S.J.J.M.F. Kokkelmans, M.J. Holland, Phys. Rev. Lett., **89**, 180401 (2002)
[127]. C.M. Varma, Phys. Rev. Lett., **61**, 2713 (1988)
[128]. A.A. Abrikosov, L.P. Gor'kov, Sov. Phys. JETP, **124**, 1243 (1961)
[129]. A.I. Larkin, Sov. Phys. JETP, **31**, 784, (1970)
[130]. A.I. Posazhennikova, M.V. Sadovskii, JETP Lett., **63**, 358 (1996)
[131]. A.V. Balatsky, I.Vechter, J.X. Zhu, Rev. Mod. Phys., **78**, 373, (2000)
[132]. M.V. Feigel'man, L.B. Ioffe, V.E. Kravtsov, E. Cuevas, Ann. Phys., **325**, 1390 (2010)
[133]. I.S. Burmistrov, I.V. Gornyi, A.D. Mirlin, Phys. Rev. Lett., **108**, 017002 (2012)
[134]. M.A. Skvortsov, M.V. Feigel'man, Sov. Phys. JETP, **117**, 487 (2013)
[135]. A. Ghosal, M.Randeria, N. Triverdi, Phys. Rev. B, **65**, 014501 (2001)
[136]. J. Baringhaus, M. Ruan, F. Edler, A. Tejeda, M. Sicot, A. Taleb-Ibrahimi, A.-P. Li, Z. Jiang, E.H. Conrad, C. Berger, C. Tegenkamp, W.A. de Heer, Nature Lett., **506**, 349 (2014)

[137]. P. Esquinazi, T.T. Heikkilä, Y.V. Lysogorskiy, D.A. Tayurskii, G.E. Volovik, JETP Lett., **100**, 336 (2014)
[138]. P.G.de Gennes, Superconductivity in metals and alloys, Benjamin, New York, 1966
[139]. M.P.A. Fisher, G. Grinstein, S.M. Girvin, Phys. Rev. Lett., **64**, 587 (1990).
[140]. A.M. Goldman, N.Markovic, Phys.Today, **51**, 39 (1998)
[141]. D.B. Haviland, Y.Liu, A.M. Goldman, Phys. Rev. Lett., **62**, 2180 (1989)
[142]. N.M. Plakida V.L. Aksenov, S.L. Drechsler, Europhys.Lett., **4**, 1309 (1987)

Figure captures

1) Fig.1 Crystal structure of graphene bilayer [11, 89]

2) Fig.2 Phase diagram of superconducting state in idealized bilayer graphene [11, 89]

3) Fig.3 $T_C$ dependence on polarization degree in 2D repulsive Fermi-gas [12, 16]

4) Fig.4 The superconducting state of 2D electron gas in a parallel magnetic field [12, 18]

5) Fig.5 Interaction potential in 2D t-J model [12]

6) Fig.6 The phase-diagram of superconducting state of 2D t-J model [12, 24]

7) Fig.7 The typical phase-diagram of BCS-BEC crossover in the resonance 3D Fermi-gas [12, 30]

8) Fig.8 Phase-diagram of the attractive-$U$ Hubbard model in 2D case [12]

9) Fig. 9 Crystal structure of superconducting alloy $BaKBiO_3$ [12, 44]

10) Fig.10 Interplay between Fermi and Bose subsystems in superconducting $BaKBiO_3$ [12, 44]